\documentclass[]{aastex631}
\usepackage{graphicx}

\newcommand{\HI}{H\,{\sc i}}

\newcommand{\kms}{\mbox{${\rm km~s^{-1}}$}}

\shorttitle{Star formation and structure of ESO006-001}
\shortauthors{Makarova et al.}

\begin{document}

\title{A nearby isolated dwarf: star formation and structure of ESO\,006-001}

\author{Lidia N. Makarova}
\affiliation{Special Astrophysical Observatory of the Russian Academy of Sciences, Nizhnij Arkhyz, Karachay-Cherkessia 369167, Russia}

\author{R. Brent Tully}
\affiliation{Institute for Astronomy, University of Hawaii, 2680 Woodlawn Drive, Honolulu, HI 96822, USA}

\author{Gagandeep S. Anand}
\affiliation{Space Telescope Science Institute, 3700 San Martin Drive, Baltimore, MD 21218, USA}

\author{Trystan S. Lambert}
\affiliation{Núcleo de Astronomía de la Facultad de Ingeniería y Ciencias, Universidad Diego Portales, Av. Ejército Libertador 441, Santiago, Chile}

\author{Margarita E. Sharina}
\affiliation{Special Astrophysical Observatory of the Russian Academy of Sciences, Nizhnij Arkhyz, Karachay-Cherkessia 369167, Russia}

\author{B\"{a}rbel S. Koribalski}
\affiliation{Australia Telescope National Facility, CSIRO Astronomy and Space Science, P.O. Box 76, Epping, NSW 1710, Australia}
\affiliation{School of Science, Western Sydney University, Locked Bag 1797, Penrith, NSW 2751, Australia}

\author{Ren\'ee C. Kraan-Korteweg}
\affiliation{Department of Astronomy, University of Cape Town, Private Bag X3, Rondebosch 7701, South Africa}

\begin{abstract}
Observations with the Hubble Space Telescope unexpectedly revealed that the dwarf galaxy ESO\,006-001 is a near neighbor to the Local Group at a distance of $2.70\pm0.11$~Mpc.  The stellar population in the galaxy is well resolved into individual stars to a limit of $M_I\sim-0.5$~mag.  The dominant population is older than 12 Gyr yet displays a significant range in metallicity of $-2 < \rm{[Fe/H]} < -1$, as evidenced by a Red Giant Branch with substantial width.  Superimposed on the dominant population are stars on the Main Sequence with ages less than 100~Myr and Helium burning Blue Loop stars with ages of several hundred Myr. ESO\,006-001 is an example of a transition dwarf; a galaxy dominated by old stars but one that has experienced limited recent star formation in a swath 
near the center.  No \HI\ gas is detected at the location of the optical galaxy in spite of the evidence for young stars.   Intriguingly, an \HI\ cloud with a similar redshift is detected 9 kpc away in projection.  Otherwise, ESO\,006-001 is a galaxy in isolation with its nearest known neighbor IC\,3104, itself a dwarf, at a distance of $\sim 500$~kpc.
\end{abstract}



\section{Introduction}
\label{sec:intro}
The nearby Universe within about 10 Mpc is a unique laboratory for studies of galaxy structure and evolution, as the galaxies in this volume are well resolved into individual stars using the Hubble Space Telescope (HST). In the last 20-25 years a number of large and successful observational programs have been carried out at HST with the Advanced Camera for Surveys (ACS) and the Wide Field and Planetary Camera 2 (WFPC2), for example as reported by \cite{2002A&A...389..812K,2006AJ....131.1361K,2009ApJS..183...67D,2013AJ....146...86T}. One goal of these programs was to ascertain the structure of the Local Universe by measuring a large number of accurate photometric distances using the Tip of the Red Giant Branch (TRGB) distance indicator \citep{1993ApJ...417..553L}. Thanks to these projects, about 500 nearby galaxies have well resolved HST/ACS (or WFPC2) images at least in two broadband filters (usually $F606W$ and $F814W$), as well as accurate TRGB photometric distances \citep{2021AJ....162...80A}. Distinct stellar populations and features are resolved in great details in the target galaxies. These extraordinary data allow us to obtain a quantitative star formation history of a nearby galaxy even from one-orbit HST observations. Great strides in our understanding have been made in the detailed nature of what should be typical galaxies and their environment \citep{2015ApJ...802...66M, 2018MNRAS.479.4136K, 2019ApJ...872...80C}.

ESO\,006-001 is one of the nearest galaxies outside the Local Group but has largely escaped attention until recently \citep{2021AJ....162...80A}.  Its existence was recorded as an entry in the ESO-Uppsala catalog of southern galaxies \citep{1982euse.book.....L} and was the subject of photometry by \citet{2002A&A...388...29P} but the combination of an erroneous redshift assignment of 738~\kms\ and the $5^\circ$ proximity of the galaxy to the difficult to observe south celestial pole contributed to its neglect. Yet the relatively low redshift sufficed to have it included as a target of imaging observations with HST.  It was a great surprise that the galaxy resolved into stars to almost 4 magnitudes below the tip of the red giant branch in a single orbit HST observation in two bands.  At the implicit close distance, if the redshift were correct the galaxy would have a large anomalous peculiar velocity away from us.  With the revised, lower redshift that we will report, the galaxy has a moderate peculiar velocity toward us.  The galaxy retains interest as an example of an isolated galaxy with moderate ongoing star formation that is close enough to be studied in detail.


\begin{figure*}
\centering
\includegraphics[width=18cm]{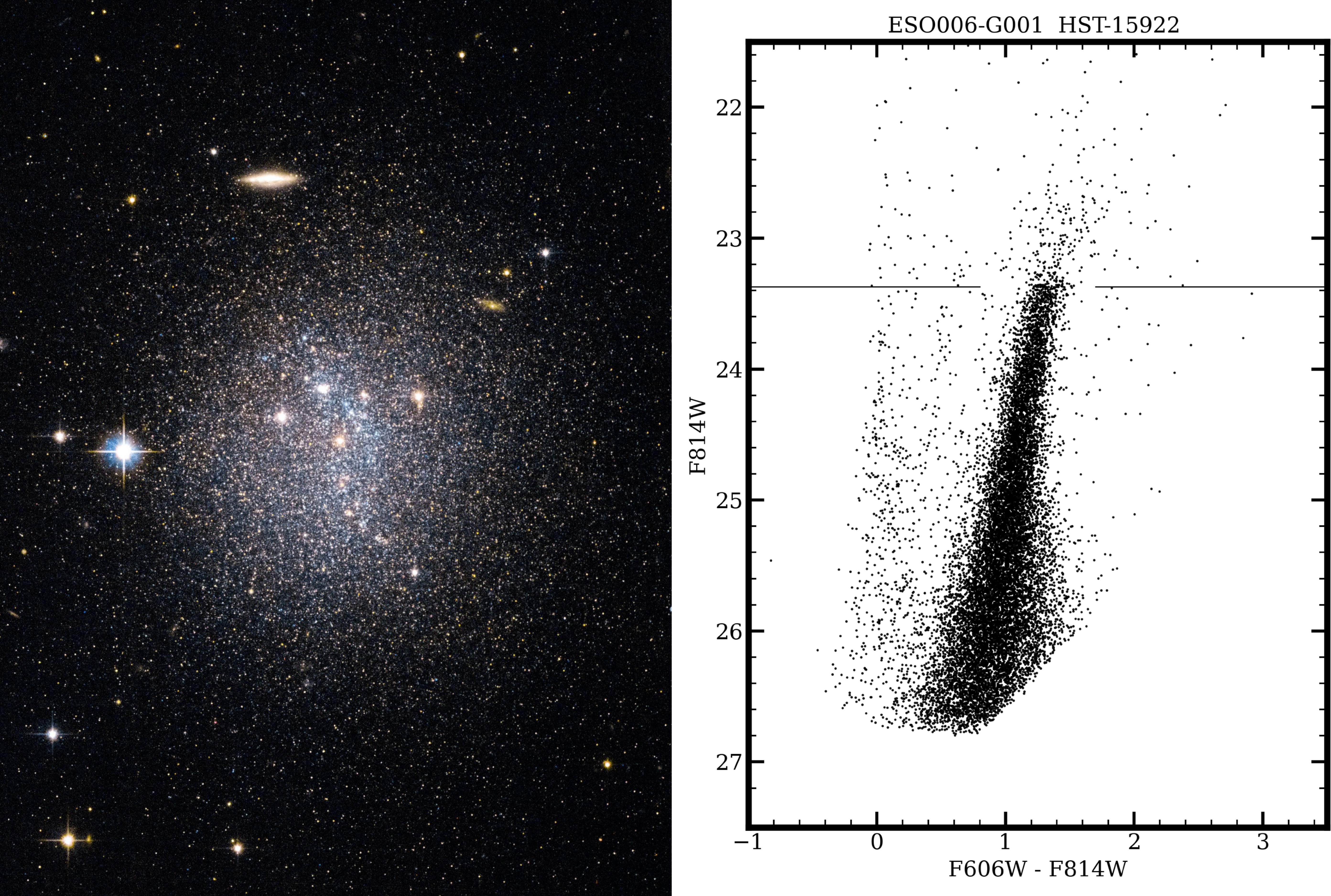}
\caption{\textit{Left:} Pseudo-color image of ESO 006-001, generated from our HST ACS/WFC imaging in $F606W$ and $F814W$ pass bands. The image size is 85 arcsec $\times$ 115 arcsec \textit{Right:} Color-magnitude diagram of resolved stars in ESO\,006-001 based on the same HST observations. The tip of the red giant branch lies at the magnitude located by the horizontal line.}
\label{fig:colorAndTRGB}
\end{figure*}

\section{HST Imaging and Galaxy Distance}
\label{sec:hstobs}

\begin{figure*}
\centering
\includegraphics[width=8cm]{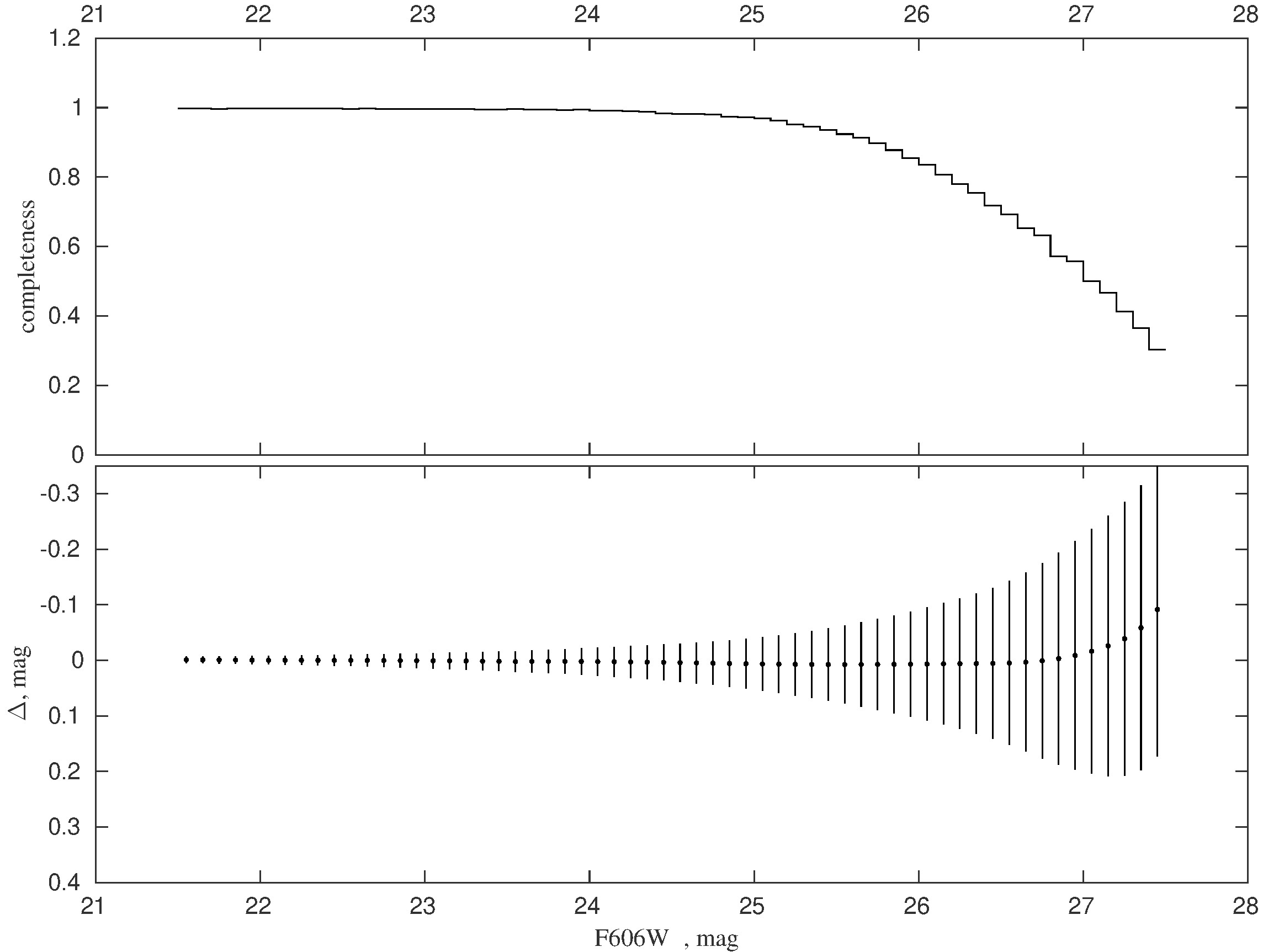}
\includegraphics[width=8cm]{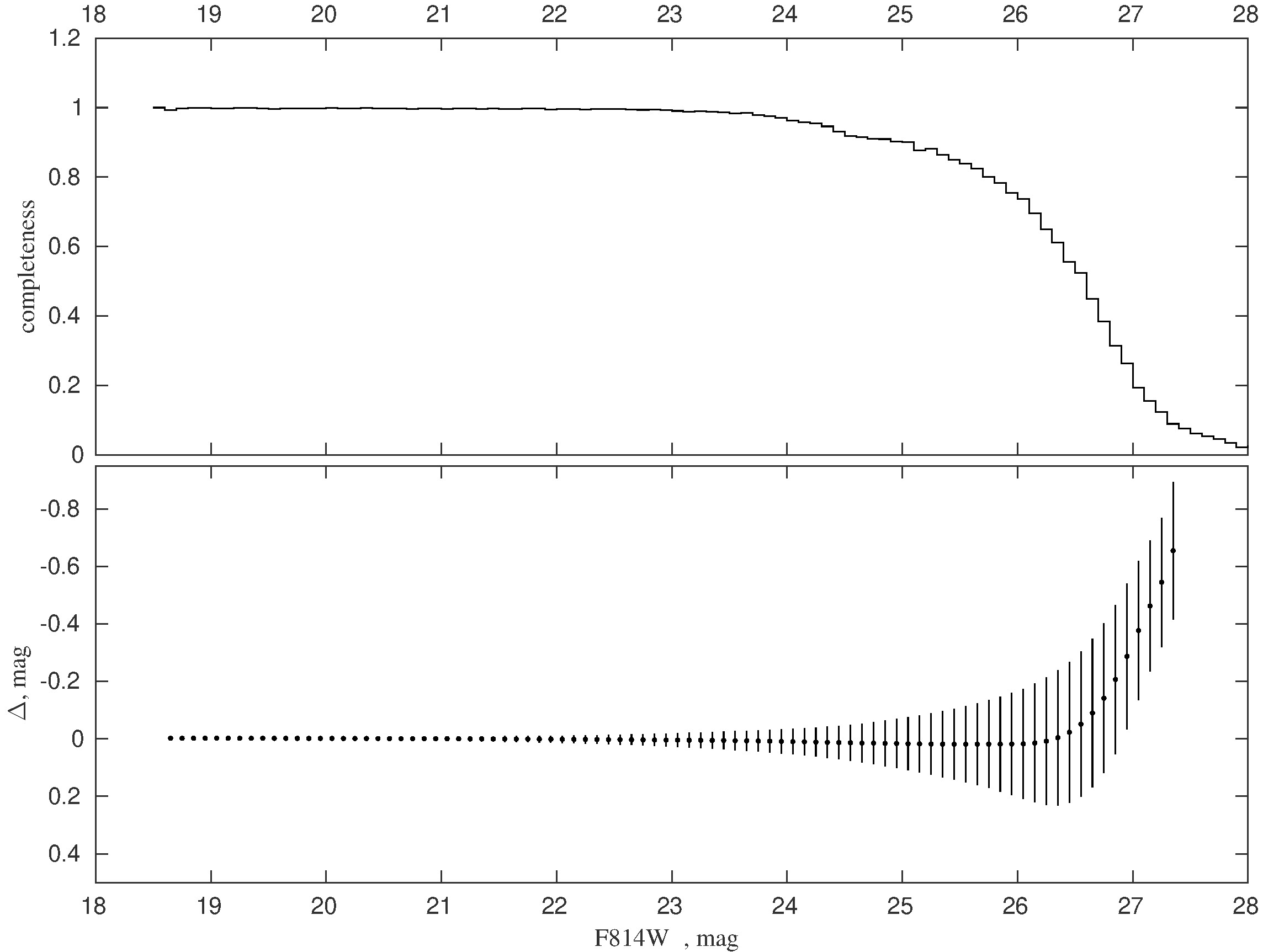}
\caption{Photometric errors and completeness values from artificial star tests. The top panels show the completeness, i.e. the fraction of artificial stars recovered within the photometric reduction procedure, as a function of the F606W and F814W magnitudes. The bottom panels give the difference between the measured and input magnitude ($\Delta$mag). The error bars are 1$\sigma$ residuals.}
\label{fig:completeness}
\end{figure*}

The dwarf galaxy ESO\,006-001 was observed as a potential object within 10~Mpc in the recently completed HST/ACS SNAP program ``Every Known Nearby Galaxy'' (PI R. B. Tully), and briefly presented as an object of interest for future studies by \cite{2021AJ....162...80A}. In this section, we recap and expand upon the data reduction presented in that work. Two exposures of 760 sec were made in each of the broadband $F606W$ and $F814W$ filters. We make use of individual \textit{flc} images obtained from MAST, calibrated through the STScI pipeline and corrected for losses in charge transfer efficiencies. For the photometry and subsequent TRGB distance measurement, we follow the methods utilized to build the CMDs/TRGB catalog in the Extragalactic Distance Database \citep{2009AJ....138..332J, 2021AJ....162...80A}. 

Briefly, we perform PSF photometry using the ACS module of DOLPHOT \citep{2000PASP..112.1383D, 2016ascl.soft08013D} using the recommended selection parameters. We use the drizzled $F814W$ image as a reference frame for alignment and source detection, with the photometry itself being performed on the underlying \textit{flc} images. The photometry is trimmed to remove sources of poor quality, following the cuts adopted by \cite{2017AJ....154...51M}. We then perform artificial star experiments with the same parameters to directly probe the photometric errors and completeness in our real photometric catalog. 
Five hundred thousand artificial stars were generated for statistically sufficient experiments. The magnitude and color range were selected the same as in the observed CMD, and the spatial distribution of the artificial stars at the ACS field reflects the distribution of the real stars within the galaxy.
With our photometry and artificial star experiments in hand, we proceed to measure the TRGB, which marks the onset of helium burning within low-mass stars, and has been increasingly used for its standard candle nature to measure distances to nearby galaxies \citep{2018ApJ...861..104H, 2019ApJ...872...80C, 2021MNRAS.501.3621A}.

 Following the methods of \cite{2006AJ....132.2729M} and \cite{2014AJ....148....7W}, we fit a model luminosity function to the observed populations of asymptotic giant branch (AGB) and red giant branch (RGB) stars, with the break denoting the location of the TRGB. As the main body of the dwarf galaxy does not take up the full ACS field of view, we limit the color-magnitude diagram (CMD) to a restricted region around the galaxy. During our procedure, we explicitly take into account the levels of photometric completeness and errors into our model luminosity function fitting procedure.
 See the right panel of Fig.~\ref{fig:colorAndTRGB} for the CMD that is derived. Photometric errors and completeness are represented in Fig.~\ref{fig:completeness}.

 The absolute calibration of the TRGB in $F814W$ has a modest dependence on metallicity and age, which is projected onto the observable color. We use the absolute and color-calibration provided by \cite{2007ApJ...661..815R}, making corrections for foreground extinction, $E(B-V) = 0.169$~mag, measured by \cite{2011ApJ...737..103S}. Our measurement of an apparent $m_{TRGB} =$ 23.37 $\pm$ 0.02~mag, when combined with an absolute $M_{TRGB} = -4.08$~mag from \cite{2007ApJ...661..815R} and $A_{F814W} = 0.292$~mag from \cite{2011ApJ...737..103S} gives a distance modulus value of $\mu = 27.16 \pm 0.09$~mag, or a distance $d = 2.70 \pm 0.11$~Mpc. The left-hand side of Figure~\ref{fig:colorAndTRGB} shows a pseudo-color image of the target, and the measured level of the TRGB is shown imposed on the CMD plot on the right-hand side of the same figure. The full-field photometric catalog, as well as a list of underlying parameters pertaining to ESO\,006-001 = PGC\,23344 can be retrieved directly from the CMDs/TRGB catalog\footnote{\url{edd.ifa.hawaii.edu}}.

\section{Globular cluster in the ESO 006-001 dwarf}
\begin{figure*}
\centering
\includegraphics[width=8cm]{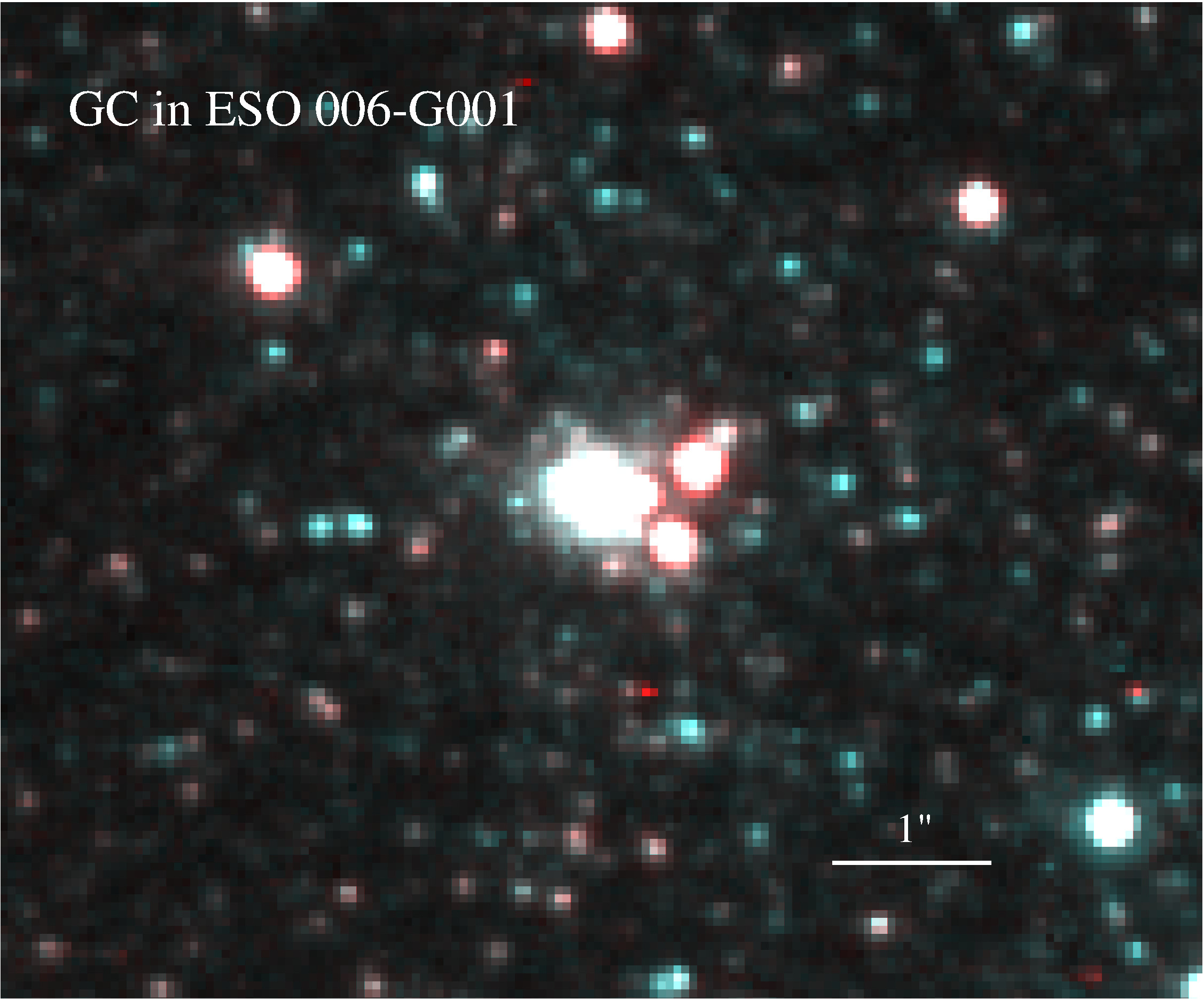}
\caption{Pseudo-color image of the found globular cluster (North is to the top, and East is to the left).}
\label{fig:gc}
\end{figure*}

A bright globular cluster with coordinates RA(J2000.0)= 08$^h$19$^m$25.012$^s$, DEC(J2000.0)=$-$85$^{\circ}$08$^{\prime}$29.15$^{\prime\prime}$ (see Fig.~\ref{fig:gc}) was found in the northern part of the elongated strip of blue stars at the HST/ACS images. Surface photometry of the cluster was performed using the method described in Section~4. According to the measured photometric parameters, the object resembles the characteristics of a globular cluster in the dwarf galaxy Sextans~B \citep{2007AstBu..62..209S}, but it looks somewhat more blue and compact.
The respective absolute magnitude, color and central surface brightness (Johnson Cousins photometric system) are $M_V{_0} = -7.4\pm0.1$~mag, $(V-I)_0 = 0.56\pm0.2$~mag and $SB_c{_0}=17.53\pm0.05$~mag. The measured half-light radius is equal to about 2~pc. The magnitudes were corrected for the Galactic extinction from \cite{2011ApJ...737..103S}. Photometry and measurements of the cluster's structural parameters are complicated by the proximity of bright red foreground stars on the western edge of the object. These stars were masked to perform surface photometry of the globular cluster.


\begin{figure*}
    \centering
    \includegraphics[height=15cm, angle=-90]{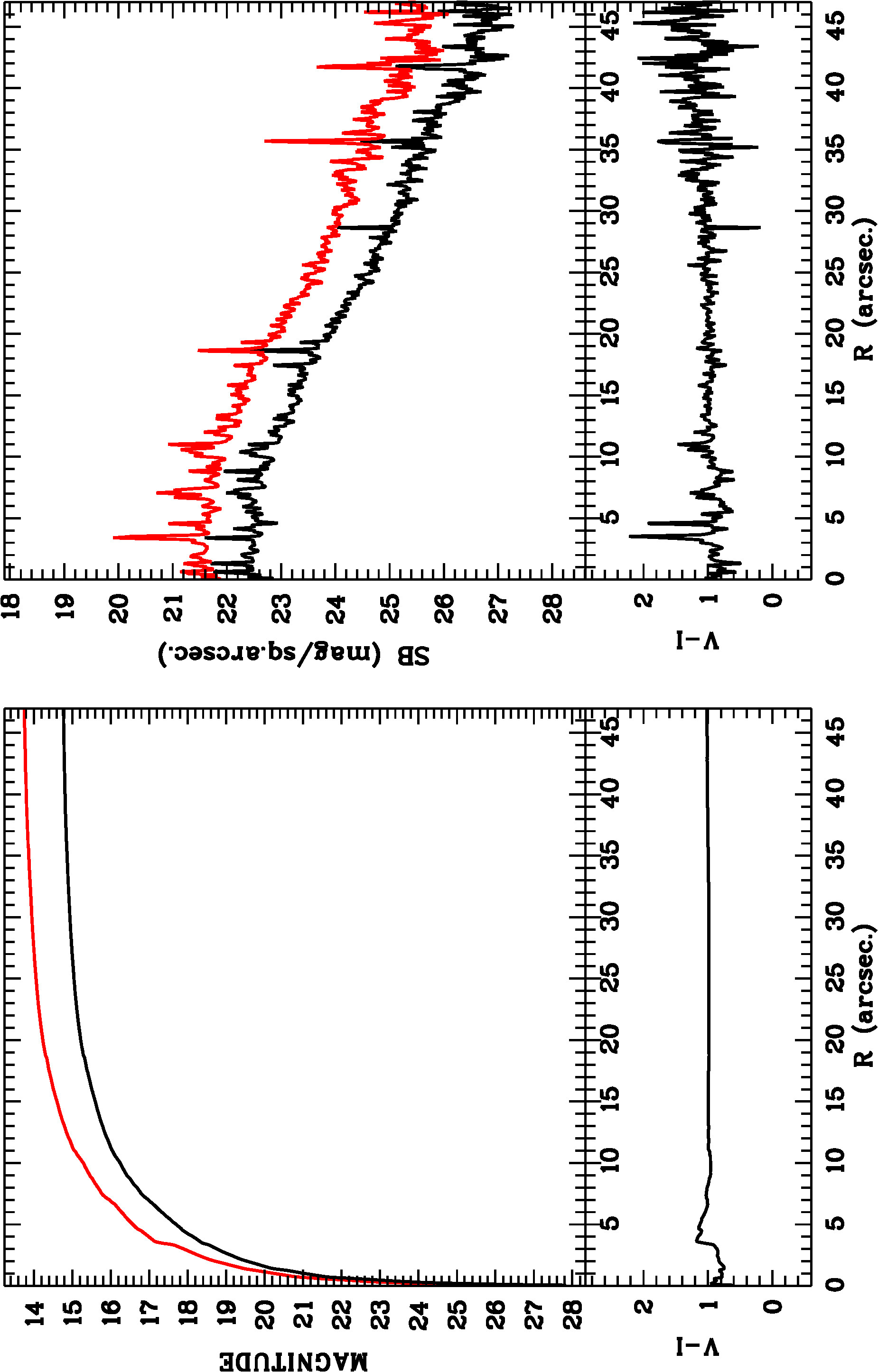}
    \caption{Integrated magnitudes and surface brightness profiles of ESO\,006-001 in the HST $F606W$ and $F814W$ bands.}
    \label{fig:sb_prof}
\end{figure*}

\begin{table*}
\begin{center}
\caption{Surface photometry results of the ESO\,006-001}
\label{table:sphot}
\begin{tabular}{lcccccr}
\hline
$V_T$    & $\rm SB^{obs}_{V_0}$& $\rm SB^{mod}_{V_0}$&  $\rm h_V$     & $\rm Re_V$ &  $\rm SB_{V_e}$&  $\rm V_e$ \\
$I_T$     & $\rm SB^{obs}_{I_0}$& $\rm SB^{mod}_{I_0}$&  $\rm h_I$     & $\rm Re_I$ &  $\rm SB_{I_e}$&  $\rm I_e$ \\
   (1)&    (2)              &     (3)             &    (4)         &    (5) &   (6)          &    (7)     \\ \hline  
14.75$\pm$0.11 &  22.35$\pm$0.08     &   22.09$\pm$0.04    &  11.08$\pm$0.2 & 17$\pm$1.5 & 22.85$\pm$0.11 &  15.46$\pm$0.05 \\
13.74$\pm$0.08 &  21.43$\pm$0.06     &   21.18$\pm$0.04    &  12.01$\pm$0.3 & 18$\pm$2.0 & 21.91$\pm$0.09 &  14.39$\pm$0.10 \\
\hline                                                               \end{tabular}                                                               \end{center}                                                                \end{table*}


\section{Surface Photometry}
\label{sec:surface_photometry}
Surface photometry of ESO\,006-001 was carried out in a manner described by \cite{2008MNRAS.384.1544S}, \cite{2013AN....334..773S} and \cite{2019Ap.....62....9S}. We used {\it SURFPHOT} from the {\it MIDAS} (Munich Image Data Analysis System) software package for astronomical data analysis \citep{1983Msngr..31...26B}. Total and surface photometry of the dwarf galaxy was made on the fully processed distortion-corrected HST/ACS $F606W$ and $F814W$ images.

The {\it FIT/BACKGROUND} program was used to fit the sky background with a tilted plane and to subtract it from the original image. The accuracy
of the sky background determination is about $1-2\%$ of the sky level. A search to define the galaxy center and a modeling of the intensity distribution over the area of the object was carried out with the ellipse fitting {\it FIT/ELL3} procedure. Then the flux was integrated and the azimuthally averaged surface brightness was calculated in concentric elliptical apertures. Additionally, integrated photometry was performed in concentric rings about the same galaxy center with steps of one pixel. The photometric results were converted from the HST/ACS photometric system to the standard Johnson-Cousins photometric system using the transformation equations from \cite{2005PASP..117.1049S}. Results are reported in Table~\ref{table:sphot} which contains the following columns: (1) - apparent $V$ and $I$ integrated magnitudes; (2) - observed central surface brightness (SB) in $V$ and $I$ bands and corresponding errors in mag\,arcsec$^{-2}$; (3) - exponential model central SB in $V$ and $I$ bands and respective errors in mag\,arcsec$^{-2}$; (4) - exponential model scale length in arcsec in $V$ and $I$ bands; (5) - effective radius in arcsec in $V$ and $I$  contained half of the total luminosity of the galaxy; (6) - respective $V$ and $I$ effective SB (mag\,arcsec$^{-2}$); (7) - effective integrated magnitudes.  Throughout, magnitudes have not been corrected for extinction.

The surface brightness profile of ESO\,006-001 (see Fig.~\ref{fig:sb_prof}) is well described by an exponential law. The galaxy is slightly more compact and has a higher effective and central surface brightness than an average Local Volume dwarf galaxies at its luminosity. We note that the photometric parameters of ESO\,006-001 are closer to those of irregular dwarf galaxies in the Virgo cluster on average than for isolated dwarf galaxies \citep{2019Ap.....62....9S}. 

ESO\,006-001 was detected with the Galaxy Evolution Explorer (GALEX) in the NUV band, and it was not observed in the FUV band. This ultraviolet emission is a direct and independent indication of recent star formation. The GALEX catalogued NUV-magnitude is 17.60 $\pm$ 0.04 (see Table~\ref{table:general}). The ultraviolet emission has an elongated shape, apparently corresponding to the narrow strip of bright blue stars that is seen in the optical image (see Sect.~\ref{sec:cmd}).

\section{Spectroscopy}
\label{sec:spectra}

Spectroscopy of ESO\,006-001 was carried out using the SpUpNIC spectrograph on the 1.9m telescope at the South African Astronomical Observatory (SAAO). This extremely low surface brightness object was
observed twice, once on the 4th of February 2021 and again on the 7th of
February 2021. The spectra were reduced following the procedure
described by \cite{huchra2012} using IRAF\footnote{IRAF is distributed
by the National Optical Astronomy Observatory, which is operated by
the Association of the Universities for Research in Astronomy (AURA) under cooperative agreement with the National Science Foundation.}. Flat-fielding and bias corrections were carried out using the \texttt{CCDRED} 
tasks following the same pipeline used by \cite{Macri2019} who employed the same instrumental setup. The spectrum was then extracted and
wavelength calibrated using the routines \texttt{APEXTRACT} and \texttt{ONEDSPEC}, respectively. The redshift of the galaxy was determined using the cross-correlation technique that involves comparing
the wavelength extracted spectrum to a series of templates using the \texttt{XCSAO} package. The templates that were used are available from the Center for Astrophysics\footnote{\url{http://tdc-www.harvard.edu/iraf/rvsao/templates/}}. The analysis resulted in a measurement of a radial velocity of $v_{\rm hel} = 319 \pm 57$~\kms.


\begin{figure}
    \includegraphics[width=8.5cm]{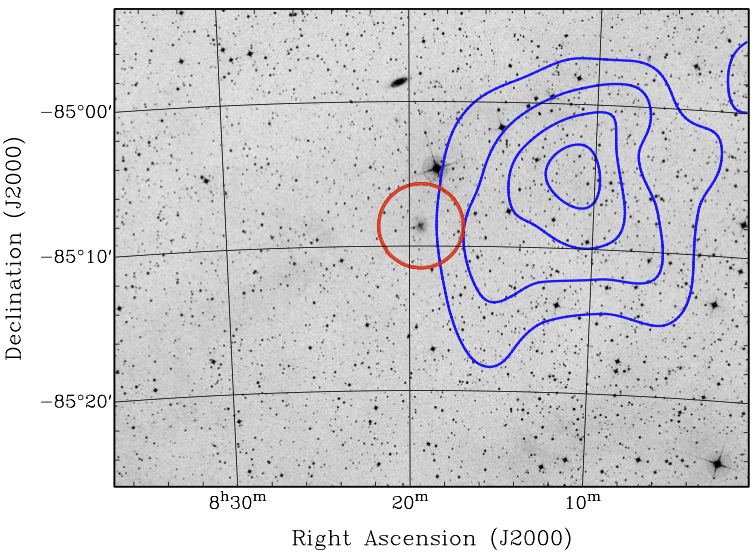}
    \caption{The dwarf galaxy ESO\,006-001 ($v_{\rm hel} = 319 \pm 57$~\kms) lying within the red circle is located in the vicinity of a compact \HI\ cloud detected in HIPASS at $333 \pm 4$~\kms. The projected separation of the peak of the \HI\ source and the galaxy is $11^{\prime} \sim 9$~kpc at the distance of the galaxy. The \HI\ contours (in blue) are 0.2, 0.6, 0.9, and 1.2 Jy\,beam$^{-1}$~\kms, slightly more extended than the $15.^{\prime}5$ beamwidth.
    }
    \label{fig:HIfield}
\end{figure}

\section{Neutral Hydrogen}
\label{sec:hi}

We analysed data from the \HI\ Parkes All Sky Survey \citep[HIPASS,][]{2001MNRAS.322..486B} which have spatial and spectral resolutions of 15.5 arcmin and 18~\kms. We establish only an upper limit to flux at the position of the galaxy but find a compact \HI\ cloud 11 arcmin to the west with $v_{\rm hel} = 333\pm4$~\kms, coincident within uncertainties with the optical velocity of ESO\,006-001.  The proximity of the \HI\ cloud to the galaxy is seen in Figure~\ref{fig:HIfield}.  The flux associated with the \HI\ cloud is $F_{\rm HI}$ = 0.75 Jy \kms\ corresponding to an \HI\ mass $M_{\rm HI}$ = $2.36~10^5 F_{\rm HI}~d^2$ = $1.3~10^6 M_{\odot}$ at the distance of the galaxy. The flux has substantial uncertainty because of confusion in the field from Galactic \HI.


\section{Cosmography}
\label{sec:xyz}

\begin{figure}
    \centering
    \includegraphics[width=7cm]{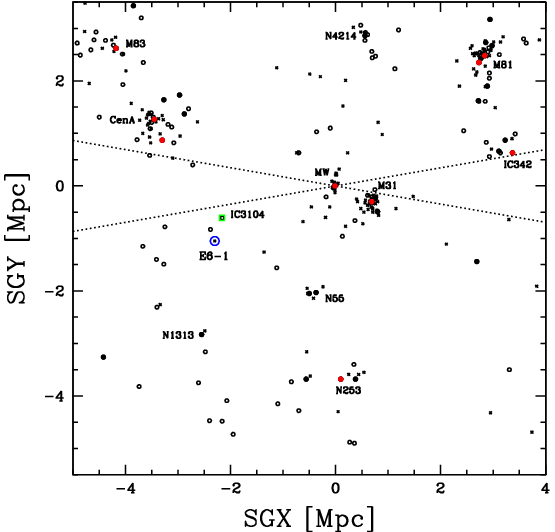}
    \includegraphics[width=7cm]{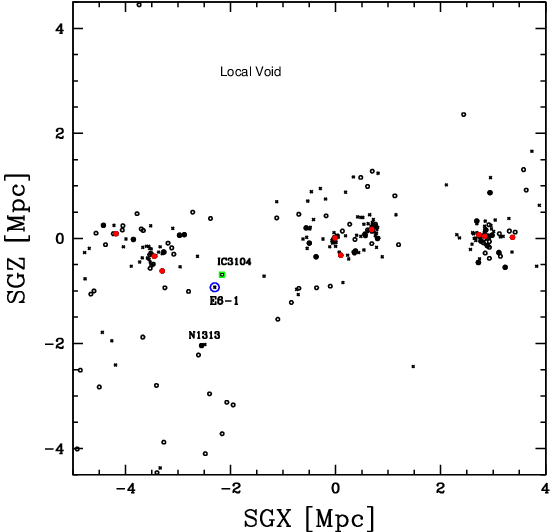}
    \includegraphics[width=7cm]{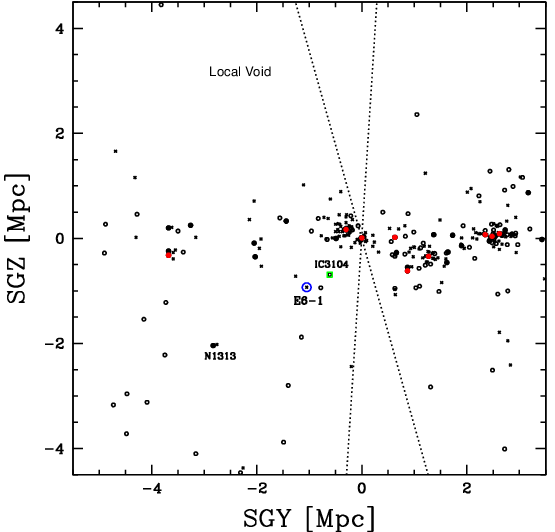}
    \caption{Three projections illustrating the distribution of nearby galaxies. ESO\,006-001 is highlighted within a blue circle.  Its nearest neighbor IC\,3104 is highlighted in green. Galaxies represented by filled red dots: log$L_K>10$; filled black dots: log$L_K>9$; small open black circles: $M_B<-13$; small crosses: dwarfs with $M_B>-13$.  Wedges opening from the origin lie at $\pm10^{\circ}$ from the Galactic plane.}
    \label{fig:xyz}
\end{figure}

The accurate distance and redshift established for ESO\,006-001 clearly establish the disposition of this galaxy with respect to the positions and motions of other nearby galaxies. Figure~\ref{fig:xyz} illustrates the distribution of galaxies within a cube 9~Mpc on a side, slightly shifted from our location at the origin in supergalactic coordinates.  The census of galaxies contributing to this plot is essentially complete outside the zone of obscuration for systems brighter than $M_B=-13$ and essentially all distances are established to within 5\% with TRGB measurements.

ESO\,006-001 finds itself within the supergalactic equatorial band, with $SGZ=-0.93$~Mpc.  From its perspective, the nearest significant groups are the Centaurus\,A and Local groups at slightly over 2~Mpc.  Its nearest neighbor is IC\,3104, itself a dwarf although somewhat larger with $M_V\simeq-13.6$~mag (roughly a magnitude brighter than ESO\,006-001 at $M_V=-12.4$).  The separation between these galaxies is $0.52^{+0.18}_{-0.13}$~Mpc. The lower limit to their separation of 0.39~Mpc is set by their separation in projection of $8.88^{\circ}$.  IC\,3104, given its luminosity, has a 3D second turnaround (splashback) radius characterizing its halo of $\sim80$~kpc surrounded by a first turnaround radius marking decoupling from cosmic expansion at $\sim240$~kpc \citep{2015AJ....149...54T}.  The corresponding values for the smaller ESO\,006-001 are $\sim 65$ and $\sim200$ kpc, respectively. Each of these two galaxies must inhabit separate halos.  There is a remote possibility that they lie within a common infall domain.

The observed velocities of ESO\,006-001 and IC\,3104 are $319\pm57$~\kms\ and $429\pm3$~\kms, respectively. The latter results from an \HI\ detection (\cite{2004AJ....128...16K} (HIPASS BGC) and \cite{2018MNRAS.478.1611K} (LVHIS), see images here: https://www.atnf.csiro.au/research/LVHIS/data/LVHIS038.info.html),
hence the small uncertainty.  These velocities translate in the Local Sheet rest frame \citep{2008ApJ...676..184T} to $V_{\rm ls}=58$ and 160~\kms, respectively. Deviations from cosmic expansion, $V_{\rm pec}$, can be calculated from distances, $d$, and assuming Hubble Constant $H_0=75$~\kms\,Mpc:
\begin{equation}
V_{\rm pec} = V_{\rm ls} - H_0d
\label{eq:vpec}
\end{equation}
whence $V_{\rm pec} = -144\pm58$~\kms\ for ESO\,006-001.  For IC\,3104, $V_{\rm pec}=-16\pm10$~\kms, where the error reflects a reasonable range of $H_0$ possibilities.  The negative peculiar velocity for ESO\,006-001 follows the pattern for galaxies at negative $SGZ$ to systematically be approaching the supergalactic equatorial plane \citep{2019ApJ...880...52A}. 

The conclusions from this brief cosmographic overview are that ESO\,006-001 lies within the Local Sheet bounding the Local Void but in isolation of influences from other galaxies.


\begin{figure}
    \centering
    \includegraphics[width=8cm]{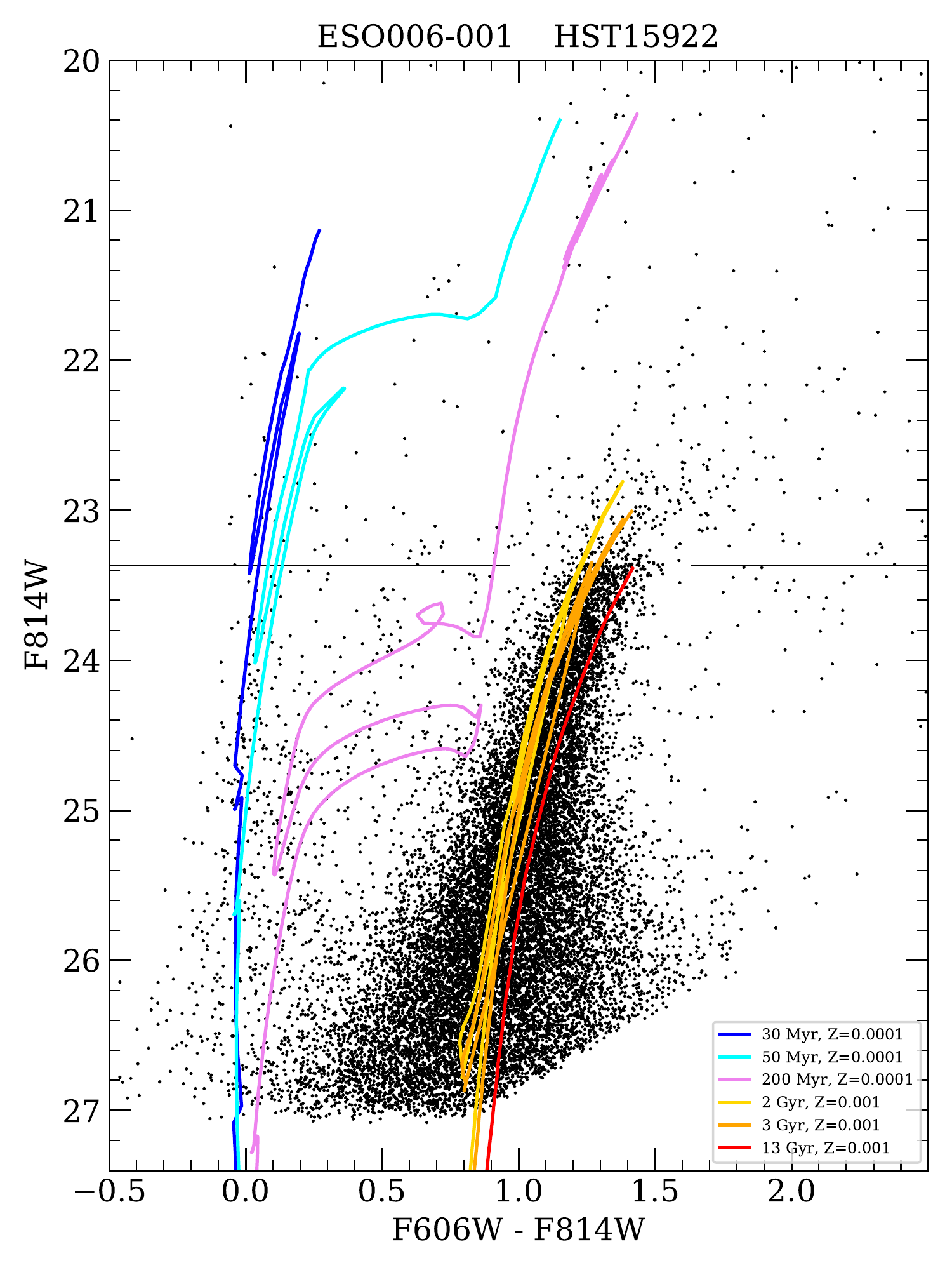}
    \caption{The $F606W-F814W$, $F814W$ color-magnitude diagram of the nearby dwarf ESO\,006-001. Magnitudes are not corrected for Galactic extinction. Black horizontal line marks the position of the TRGB at $F814W=23.37$ mag. PARSEC stellar isochrones are overlaid \citep{bressan2012, marigo2013}. }
    \label{fig:cmd}
\end{figure}

\begin{figure}
    \centering
    \includegraphics[width=8cm]{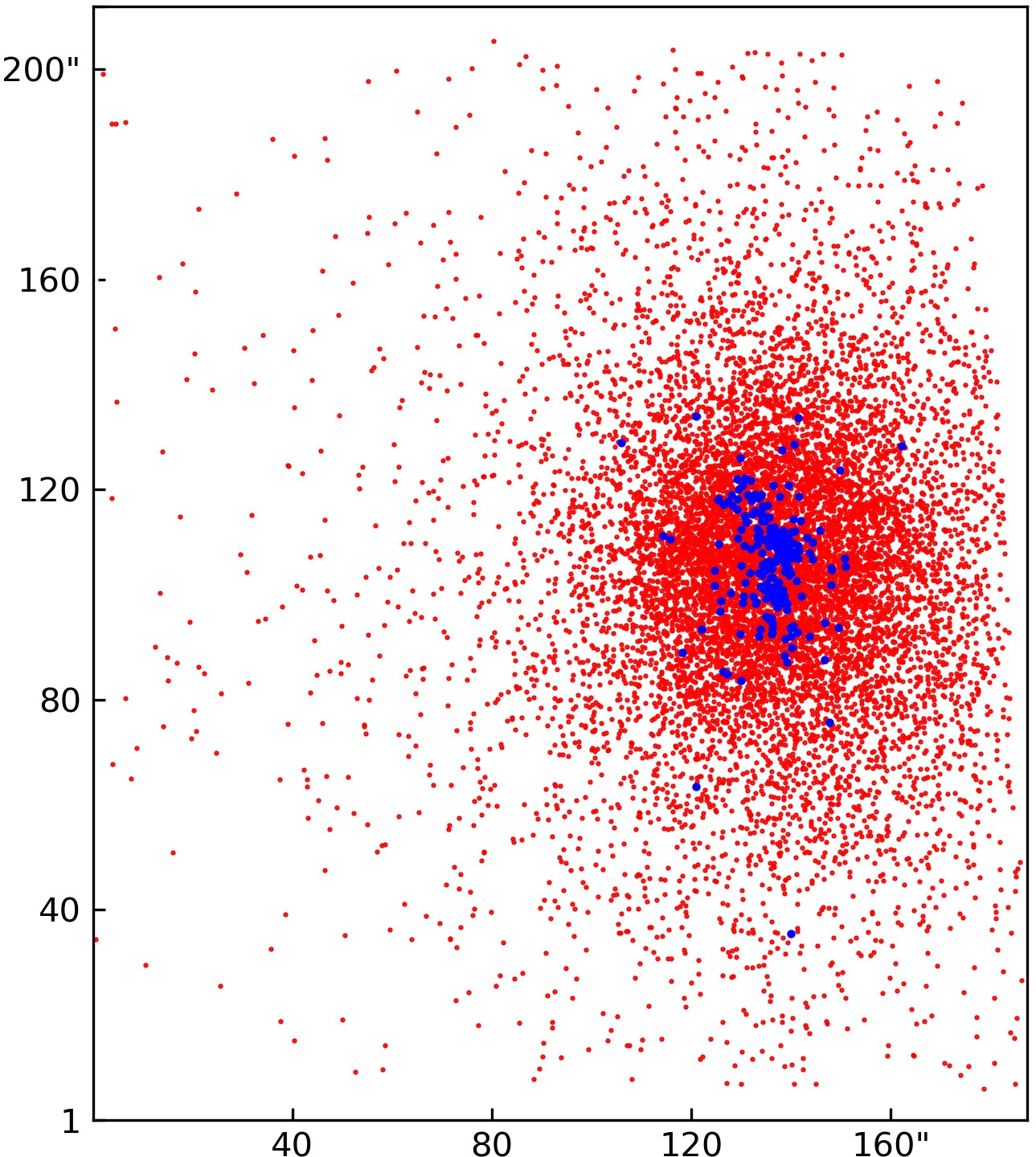}
    \caption{Spatial distribution of the different stellar populations within the body of the ESO\,006-001 galaxy in the HST/ACS image. Red dots represent RGB stars, and blue ones represent stars on the upper part of the main sequence.}
    \label{fig:spatial}
\end{figure}

\section{Color-magnitude diagram overview}
\label{sec:cmd}

The dwarf galaxy ESO\,006-001 is well resolved into individual stars and
a number of qualitative features are discerned in its color-magnitude diagram seen in Fig.~\ref{fig:cmd}. Deep photometry of the ACS field indicates the presence of resolved stars associated with the galaxy extending to the edges of
the ACS field (see Fig.~\ref{fig:spatial}). Therefore, for the detailed star formation history and the color-magnitude diagram analysis we used the entire ACS field.
In the $F606W-F814W$ color interval between $-0.5$ and $0.5$ mag,
the upper main sequence (MS) is sparsely populated but plausibly extends as bright as $F814W\sim 20$~mag ($M_I\sim -7$~mag).
A number of probable blue loop stars can be distinguished between 
$F606W-F814W\sim0.3$ and $\sim 0.8$~mag and fainter than $F814W\sim21.2$~mag.
The sparsely populated red supergiant branch (RSG) and upper part of the asymptotic giant
branch (AGB) are visible redder than $F606W-F814W\simeq1.0$~mag and brighter
than $F814W$ $\sim22$~mag. The lower part of AGB is unambiguously distinguished but not prominent between
$F814W\sim22$~mag and the tip of the red giant branch (TRGB) at
$F814W=23.4$~mag.
The most conspicuous and highly populated feature in the CMD is the red giant
branch (RGB) situated with $F606W-F814W$ between 0.7 and 1.5 mag and lower than $F814W=23.4$ mag.

Fig.~\ref{fig:spatial} shows the spatial distribution of populations of different ages within the galaxy. Upper main sequence stars were selected in the color range $F606W-F814W = -0.5$ to 0.4 mag 
and in the magnitude range $F814W = 25.0$ to 20.0 mag. These stars are identified by blue dots in
Fig.~\ref{fig:spatial}. Main Sequence stars reflect recent/ongoing star formation, currently taking place in a narrow, elongated strip across most of the galaxy. RGB stars are indicated in the Fig.~\ref{fig:spatial} in red, and were selected in the appropriate range $F606W-F814W = 0.6$ to 2.0 mag and $F814W = 26.0$ to 23.4 mag.
The old RGB stars are distributed smoothly, with a roughly exponential concentration towards the geometric center.


\begin{figure*}
    \centering
    \includegraphics[width=15cm]{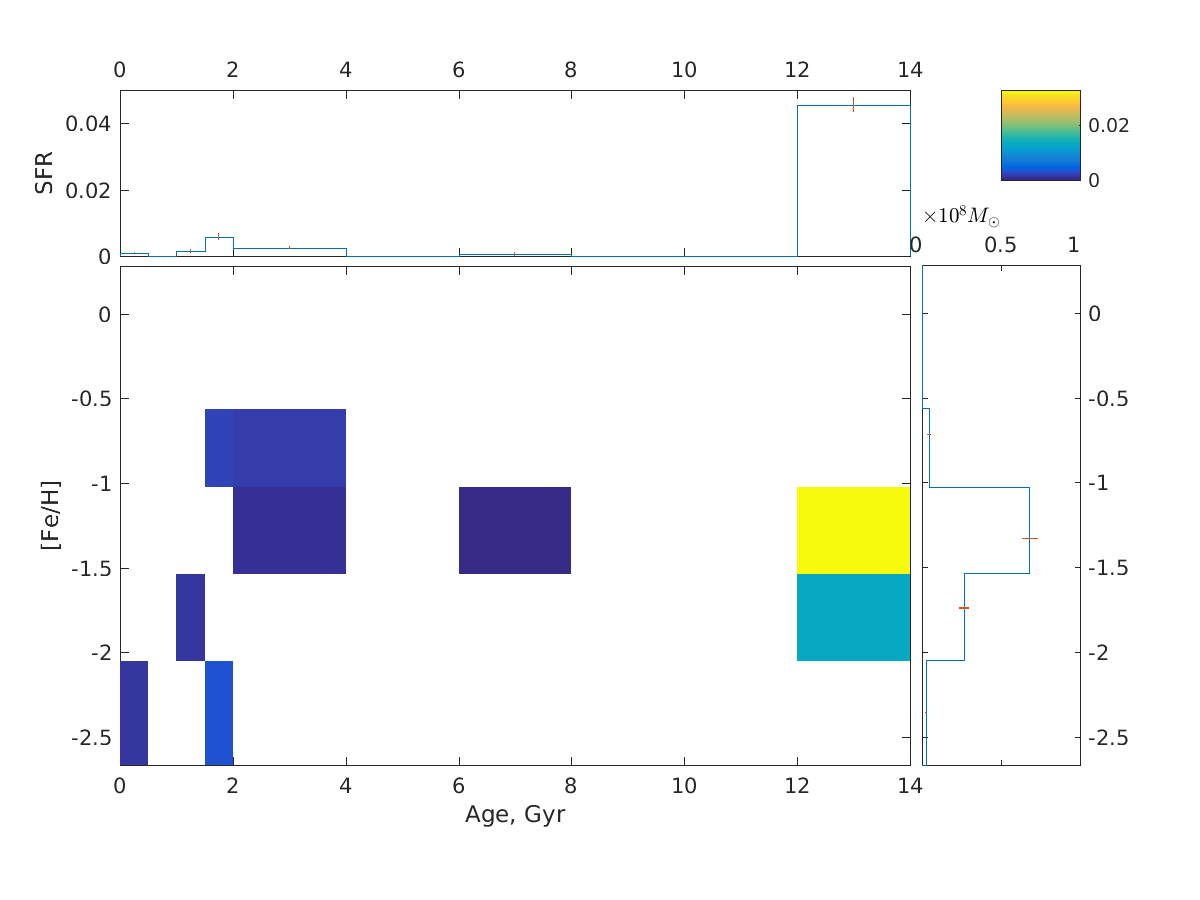}
    \caption{Results of the star formation history measurements of
    of the ESO\,006-001 dwarf galaxy. {\it Top panel:} star formation rate (SFR) ($M_\odot$/yr) against the age of the stellar populations. {\it Bottom panel:} metallicity of the stellar content as the function of age. The colored boxes give ranges of metallicity vertically in horizontal spans of age. {\it Right panel:} the estimated stellar mass vs. metallicity. The formal errors in SFR are indicated with the vertical bars. The error bars are $1 \sigma$ uncertainties of the maximum likelihood estimation.}
    \label{fig:sfh}
\end{figure*}

\section{Star formation history}
\label{sec:sfh}

We made our first attempt at star formation history modelling
in the work of \cite{2021AJ....162...80A}. Here we study the star formation history of this nearby dwarf in more detail, with updated underlying isochrones and a more sophisticated, maximum likelihood methodology. The quantitative star formation history (SFH) of ESO\,006-001 was evaluated
using our StarProbe software package \citep{2004Ap.....47..229M}. This software and method were used and described in some details in a number of our studies of resolved dwarf galaxies in the Local Universe (see, for example, \cite{2010MNRAS.406.1152M,2017MNRAS.464.2281M,2021MNRAS.502.1623M} and reference therein). This program
determines a star formation rate (SFR) and metallicity of a galaxy with
respect to the stellar population age. We use theoretical stellar isochrones to model the resolved stellar populations. The star formation history of the galaxy is estimated by comparison of the model color-magnitude diagram with the real data.
A distribution of stars in any color-magnitude diagram can be represented as a linear superposition of simple stellar populations (SSPs) of different ages. The distance to the object under study, photometric errors, Galactic and internal extinction affect this photometric distribution. An accurate photometric distance of ESO\,006-001 is known from the luminosity of the TRGB and the well established estimates of photometric errors from the extensive artificial star tests (see Sect.~\ref{sec:hstobs}).
In particular, the same incompleteness and crowding effects, and photometric systematics as those determined for the observations using artificial stars experiments were used in our modelling.
All the selected stars of good photometric quality in the CMD of ESO\,006-001 (see Sect.~\ref{sec:hstobs}) were used in the SFH determination. No additional weight is assigned to any local part the observational CMD.
The Galactic extinction values were taken from \cite{2011ApJ...737..103S}. 
For our calculation we assume that internal extinction is negligible, as in small, low-metallicity dwarf galaxies the dust content is usually low (see, for example, \cite{2011A&A...529A.149G, 2014A&A...563A..27L}).

 We construct a model color magnitude (Hess) diagram from theoretical stellar isochrones. Each isochrone corresponds to 
a specific age and metallicity, and the entire set of these models fits a fairly wide range of ages (from 0 Myr to 14 Gyr)  and 
metallicities (from Z = 0.0001 to Z = 0.03) of stellar populations,
with an age resolution 2 Gyr steps for the older populations and 0.5 Gyr steps near 2 Gyr and younger. The isochrones were interpolated in age, to avoid discontinuities, so that the sampled points in the CMD are separated by at maximum 0.03 mag. A metallicity resolution in our models is set by the underlying published isochrones, and they were not interpolated.
We find 
an appropriate linear combination of partial model CMDs for the best approximation of the observed CMD. We construct the analytical distribution function of stars in the Hess diagram for each isochrone, taking into account an initial mass function (IMF), photometric errors, bin size of the Hess diagram, distance modulus and extinction. We use the Padova stellar isochrone calculations made with the PARSEC library\footnote{\url{http://stev.oapd.inaf.it/cgi-bin/cmd}} and the Salpeter IMF: $\rho(m)\,dm\sim m^{-2.35}dm$. The chosen normalization gives the total integral probability equal to one in the range of 0.1--100~$M_\odot$. The mass of a star determines its observational characteristics for a given isochrone. With the parameters mentioned above it determines the probability of a star to occur in any bin of the Hess diagram. We assume a constant star formation rate in a given age range for a combination of individual isochrones of the same metallicity. To determine a SFH, we firstly find the most significant variables (i.e., partial model CMDs) that differ from zero with a given probability. Then we determine values of the significant variables to get the desired solution by the maximum likelihood method.

\begin{table*}
\centering
\caption{General parameters of the ESO\,006-001 dwarf}
\label{table:general}
\begin{tabular}{lll}
\hline
              & ESO\,006-001 \\
\hline
Position (J2000)$^a$ &08$^h$19$^m$23.3$^s-$85$^{\circ}$08$^{\prime}$44$^{\prime\prime}$  \\
$E(B-V)$, mag$^b$    & 0.169  \\
$V_T$, mag & 14.75$\pm$0.11  \\
$I_T$, mag & 13.74$\pm$0.08  \\
$M_V$, mag & $-12.41\pm0.14$  \\
$M_I$, mag & $-13.42\pm0.12$  \\
Axes ratio, a/b & $1.13\pm0.013$ \\
Distance modulus, mag & 27.16 $\pm$ 0.09 \\
Distance, Mpc & 2.70 $\pm$ 0.11 \\
Radial velocity, km~s$^{-1}$ & 319 $\pm$ 57  \\
GALEX NUV, mag & 17.60 $\pm$ 0.04 \\
Mass fraction of oldest stars (12–14 Gyr), \% & $90\pm4$ \\
Mean SFR 12--14 Gyr ago, M$_{\odot}$/yr & $4.6\pm0.2\times 10^{-2}$ \\
Mean metallicity of oldest stars, [Fe/H], dex & $-1.5\pm0.2$ \\
Fraction of intermediate age stars (1--4 Gyr), \% & $\sim9\pm5$ \\
Mean SFR 1--4 Gyr ago, M$_{\odot}$/yr & $2.8\pm1.3\times 10^{-3}$ \\
Mean metallicity of intermediate age stars, [Fe/H], dex & $-1.3\pm0.5$ \\
Fraction of young stars, \% & $\sim1\pm0.1$ \\
Current SFR ($\leq$ 500 Myr), M$_{\odot}$/yr & $1.1\pm0.1\times 10^{-3}$ \\
Mean metallicity of young stars & $-2.4\pm0.3$ \\
Total stellar mass, M$_{\odot}$ & $1.0\pm0.1\times 10^8$ \\
\hline
\multicolumn{2}{l}{$^a$ LV database: https://www.sao.ru/lv/lvgdb} \\
\multicolumn{2}{l}{$^b$ \cite{2011ApJ...737..103S}}
\end{tabular}
\end{table*}

The results of our measurements of the star formation history of ESO\,006-001 are shown in Fig.~\ref{fig:sfh}. The $1 \sigma$ error of each simple stellar population is derived from the analysis of the likelihood function. According to our calculations, the first and most intensive burst of star formation occurred about 12-14 Gyr ago. In particular, we can say that the absolute majority of the stars (about 90\%) were formed in this first period of star formation. This star burst seems quite intensive for a dwarf object with the mean SFR of $4.6\pm0.2\times 10^{-2}$ $M_\odot$/yr. We also infer non-intensive star formation in the period of 1-4 Gyr ago favouring higher SFR 1.5-2 Gyr ago.
The maximum mean SFR does not exceed $3.5\pm0.7\times 10^{-3}$ $M_\odot$/yr at that time.
There are clear but weak signs of current star formation with the rather
low SFR $1.1\pm0.1\times 10^{-3}$ $M_\odot$/yr. The mean metallicity tends to be relatively high for a dwarf galaxy.
The measured total stellar mass is $1.0\pm0.1\times 10^8$ $M_\odot$.
The results of our study are summarized in Table~\ref{table:general}.

Overall, the measured SFH of the dwarf indicates a strong burst of star formation at an early epoch, when most of the stars in the galaxy were formed, then a pronounced gap at "intermediate" ages, and small but clear signs of renewed star formation at a recent epoch. This pattern is not rare in samples of isolated local dwarf galaxies. For example, in the work of \cite{2015MNRAS.450.4207B} authors found that nearby isolated dwarf galaxies show a wide diversity of SFHs, including ‘two-component’ systems characterized by star formation in early and late epochs and with a "gap" at intermediate times (4 Gyr $<$ t $<$ 8 Gyr).
\citet{2019MNRAS.482.1176W} have demonstrated in their simulations that nearly 20 per sent of isolated dwarfs cease and resume star formation at least once during their lifetime. The authors suggest that dwarfs resume star formation as a result of interactions with streams or filaments of gas originating from nearby mergers or the cosmic web.

Our observations allow us
to reach an absolute I stellar magnitude of roughly --0.5 mag, but critical fainter stellar populations (horizontal branch, red clump, main sequence turnoff) are below our photometric limit. Without the information contained in these age-sensitive CMD features it is difficult to resolve the age--metallicity--SFR relation for the oldest ($>$ 6--8 Gyr) star formation events, due to tight packing of the corresponding isochrones at the brightest part of the CMD. Nevertheless, in their promising tests \citet{2011ApJ...739....5W} have shown that a recovered SFH depends less on the photometric depth and more on the number of stars in the CMD if the stellar models that are used are known quite precisely. Unfortunately the stellar models are not always self-consistent, and the measured SFR may be systematically shifted in a particular time bin, depending on the stellar models and photometric depth of the CMD.

Taking into account the uncertainties mentioned above, we can nevertheless have confidence that ESO\,006-001 has followed the star formation trend common to a number of nearby isolated dwarfs, whereby active star formation at early epochs was suppressed and then recently resumed, presumably due to an interaction with filaments of gas in the cosmic web.



\section{Summary}
\label{sec:sum}

ESO\,006-001, determined to lie at $2.70\pm0.11$~Mpc from an accurate TRGB measurement, is one of the nearest galaxies outside the Local Group.  It had evaded serious attention because of its inconvenient location near the south celestial pole and an incorrect published redshift that did not hint of its proximity.  Even with the short exposures of a single orbit observation with a SNAP program with HST, individual stars could be resolved to as faint as $M_I \sim -0.5$~mag, the proximity of the Red Clump.

The dominant stellar populations in ESO\,006-001 are determined to be ancient, older than 12~Gyr.  Even with that great age, the metal abundances are considerable for such a small galaxy, reaching [Fe/H]$\sim-1$.  The old stars are, on the one hand, strongly concentrated while, on the other hand, semi-spherically and widely scattered.  The great age of most stars not withstanding, the galaxy has experienced star formation over the last few Gyr and particularly within the last few hundred million years.  The youngest stars lie in a $20^{\prime\prime}\sim260$~pc band across the central part of the galaxy as evident in the left panel of Fig.~\ref{fig:colorAndTRGB} and the GALEX detection. A bright globular cluster was found in the nothern part of the young stars' band. The absolute stellar magnitude and color estimated from the HST/ACS galaxy images are $M_V{_0} = -7.4\pm0.1$~mag, $(V-I)_0 = 0.56\pm0.2$~mag. The upper limit on the \HI\ flux at the position of the galaxy is surprising given the evidence for recent star formation.  ESO\,006-001 was already identified by \citet{2002A&A...388...29P} as a dwarf in transition between a gas rich irregular with ongoing star formation and a gas poor spheroidal depleted of star formation.  The apparent existence of a companion $10^6 M_{\odot}$ \HI\ gas cloud at a projected separation of 9~kpc is a curious situation that warrants further investigation.  

The environment of ESO\,006-001 is one of isolation.  The nearest known neighbor, the dwarf irregular galaxy IC\,3104, lies at a distance of $\sim500$~kpc (and no closer than the projected separation of 390~kpc), a distance too far removed to expect such a small galaxy to have any effect.  On a larger scale, ESO\,006-001 lies within the band of galaxies on the supergalactic equator, part of a structure suspected to be a wall arising from the expansion of the Local Void.  ESO\,006-001 adds one more piece to an increasingly compete understanding of our local neighborhood. 

\begin{acknowledgments}
This research is based on observations made with the NASA/ESA Hubble Space Telescope obtained from the Space Telescope Science Institute, which is operated by the Association of Universities for Research in Astronomy, Inc., under NASA contract NAS 5–26555. These observations are associated with program SNAP-15922.
Support for the program SNAP-15922
(PI R.B.Tully) was provided by NASA through a grant from the
Space Telescope Science Institute.
Some of the data presented in this paper were obtained from the Mikulski Archive for Space Telescopes (MAST) at the Space Telescope Science Institute. The specific observations analyzed can be accessed via \dataset[DOI]{https://doi.org/10.17909/xh0y-fs90}.
RCKK acknowledges the support by the South African Research
Chairs Initiative of the Department of Science and Innovation and the
National Research Foundation. We would also like to express our thanks to Hartmut Winkler and Francois van Wyk who obtained the spectral data at the SAAO. L.N.M. are supported in part by the Russian Federation grant 075-15-2022-262 (13.MNPMU.21.0003).
In this study we used related results of GALEX, which is a NASA mission managed by the Jet Propulsion Laboratory.
\end{acknowledgments}

\bibliography{esosfh}{}
\bibliographystyle{aasjournal}

\end{document}